\documentclass[twocolumn,aps,letter]{revtex4}
\usepackage{graphicx,graphics,float}% Include figure files
\usepackage{dcolumn}% Align table columns on decimal point
\usepackage{blindtext}
\usepackage{bm}% bold math
\usepackage{amsmath,amssymb}
\usepackage{verbatim}
\usepackage{mathdots}
\usepackage{mathrsfs}
\usepackage{xcolor}
\begin{document}
\title{Uncoupled Majorana fermions in open quantum systems: On the efficient simulation of non-equilibrium stationary states of quadratic Fermi models}
\author{Jose Reslen}
\affiliation{Coordinaci\'on de F\'{\i}sica, Universidad del Atl\'antico, Kil\'ometro 7 Antigua v\'{\i}a a Puerto Colombia, A.A. 1890, Barranquilla, Colombia.}%
\date{\today}

\newcommand{\Keywords}[1]{\par\noindent{\small{\em Keywords\/}: #1}}
\begin{abstract}
A decomposition of the non-equilibrium stationary state of a quadratic Fermi system
influenced by linear baths is obtained and used to establish a simulation protocol 
in terms of tensor states. The scheme is then applied to examine the occurrence of 
uncoupled Majorana fermions in Kitaev chains subject to baths on the ends. The 
resulting phase diagram is compared against the topological characterization of the
equilibrium chain and the protocol efficiency is studied with respect to this 
model. \\
\end{abstract}
\maketitle
\section{Introduction}
Nonequilibrium physics offers a more complete description of quantum structures by
taking into account the system interaction with environment components that are
more complex than thermalization baths. This research area is of fundamental
importance since the effect of dissipation can hardly be played down in a physical
model without severely compromising the study's application-scope. The complication
that usually arises when treating quantum systems subject to external forces, also
known as open quantum systems, is that the inclusion of baths enlarges the analysis
ambit, adding to the exponential growth of Hilbert spaces with respect to size.
Although a great deal of effort has been channeled into the search of efficient
simulation protocols for isolated quantum systems, research addressing simulation
strategies in open quantum systems is not as prolific.  The issue becomes relevant
in recent times as a growing interest in the topology of dissipative configurations
is noticeable in the quantum physics community. One of the findings that has
motivated this interest is the realization that the Kitaev chain \cite{kitaev}
displays a phase transition from local to topological. In the latter case the
system displays uncoupled Majorana fermions that present promising potential in the
field of quantum computation \cite{aguado,oreg}. In this context, it becomes
natural to inquire how out-of-equilibrium processes, intentionally induced or not,
influence the state, especially its most resilient excitations in the equilibrium
picture. It has been seen that in small chains this influence can be beneficial
under specific circumstances \cite{carmele}, but how robust this phenomenology is
against growing size is so far not entirely understood. Reference \cite{zunkovi}
reports the decay of correlations as a function of size in stationary states of XY
spin chains with baths on the ends over the whole spectrum of the model parameters,
the only difference being the decay functionality, which ultimately determines the
system's phase diagram.  Depending on a number of factors, correlations can also be
made to linger in dissipative systems, as for example when measured as entanglement
entropy in operator space \cite{pizorn}, in XXZ spin chains
\cite{znidaric,prozen_ansatz}, or as a function of time in XY chains \cite{kos}.
In this paper the issue of correlations in stationary states is addressed in
comparison with the topological features under equilibrium in a scenario where
dissipation breaks the symmetry sustaining the topological phase.  This task is
undertook over the Kitaev chain because its topological attributes are well
characterized and can be monitored using end-to-end correlations \cite{reslen5}. 

The Majorana chain is governed by the Hamiltonian \cite{kitaev,aguado,oreg}
\begin{gather}
\hat{H} = \sum_{j=1}^N -w ( \hat{c}^{\dagger}_j \hat{c}_{j+1} + \hat{c}^{\dagger}_{j+1} \hat{c}_j ) - \mu \left( \hat{c}^{\dagger}_j \hat{c}_{j} - \frac{1}{2} \right ) 
+ \nonumber \\ 
\Delta \hat{c}_j \hat{c}_{j+1} + \Delta^* \hat{c}^{\dagger}_{j+1} \hat{c}^{\dagger}_j.
\label{kita}
\end{gather}
Constants $w$ and $\mu$ are intensity parameters corresponding to the hopping and
chemical potential of a quantum wire. Constant $\Delta$ is the intensity of the
proximity effect generated by a p-wave superconductor. The model features a system
of spinless fermions described by ladder operators obeying $\{\hat{c}_j,\hat{c}_k\}
= 0$ and $\{\hat{c}_j,\hat{c}_k^\dagger\} = \delta_j^k$.  The chain boundary is
fixed, $\hat{c}_{N+1}=0$. By means of a Jordan-Wigner transformation the model
shifts to a Heisenberg XY-spin-chain. The system also admits a description in terms
of Majorana operators, $\hat{\gamma}_k = \hat{\gamma}_k^\dagger$, with the property
$\{ \hat{\gamma}_k, \hat{\gamma}_j \} = 2 \delta_k^j$. This allows to write the
original modes as
\begin{gather}
\hat{c}_{j} = \frac{1}{2} \left( \hat{\gamma}_{2j-1} +  i \hat{\gamma}_{2 j} \right),\text{ } \hat{c}_{j}^\dagger = \frac{1}{2} \left( \hat{\gamma}_{2j-1} -  i \hat{\gamma}_{2 j} \right), \label{jinga}
\end{gather}
and likewise the Hamiltonian
%
%
%{\scriptsize
\begin{gather}
\hat{H} = \frac{i}{2}\sum_{j=1}^N -\mu \hat{\gamma}_{2j-1} \hat{\gamma}_{2j} + (|\Delta|-w) \hat{\gamma}_{2j-1} \hat{\gamma}_{2j+2} + \nonumber \\
	(|\Delta|+w) \hat{\gamma}_{2j} \hat{\gamma}_{2j+1} = \frac{1}{2} \sum_{j=1}^{N} \sum_{k=j}^{N} A_{2 j - 1,2 k} \hat{\gamma}_{2 j - 1} (i \hat{\gamma}_{2 k}). \label{ostinato}
\end{gather}
%}
%
%
By definition $A_{j,k}=0$ if $j>k$. The chain is connected to linear baths 
that can in general be described by
\begin{gather}
\hat{L}_n = \sum_{j=1}^N B_{2j-1}^{(n)} \hat{\gamma}_{2j-1} + B_{2 j}^{(n)} i \hat{\gamma}_{2 j}. 
%\Rightarrow \hat{L}_n^\dagger = \sum_{j=1}^N B_{2j-1}^{(n)} \hat{\gamma}_{2j-1} - B_{2 j}^{(n)} i \hat{\gamma}_{2 j},
\label{baths}
\end{gather}
Coefficients $B_{j}^{(n)}$ are defined as real. The system dynamics can be studied
using the Lindblad master equation  ($\hbar = 1$) \cite{lindblad}
\begin{gather}
\frac{d \hat{\rho}}{d t} = -i[\hat{H},\hat{\rho}] + \sum_n 2 \hat{L}_n \hat{\rho} \hat{L}_n^\dagger - \{ \hat{L}_n^\dagger \hat{L}_n, \hat{\rho} \},
\label{lindblad}
\end{gather}
being $\hat{\rho}$ the system's density matrix. The Lindblad equation is a general
Markovian map that preserves the trace as well as the positivity of $\hat{\rho}$ in
a non-unitary fashion. This study focuses on the state the system evolves toward as
time goes to infinity, also known as the Non Equilibrium Stationary State (NESS).
It is known that the NESS of a fermion system described by a quadratic Hamiltonian
and subject to linear baths corresponds to a Gaussian state \cite{bravyi}. The most
direct way of finding the NESS, should it exist, is equating the rhs of
(\ref{lindblad}) to zero and algebraically solving for $\hat{\rho}$, but this
approach becomes inefficient very rapidly as $N$ grows, making it impractical to
study the big size behavior. It has been pointed out by Prosen in \cite{prozen}, in
similarity with the general notions of reference \cite{manfred}, that models like
this one admit a description in terms of a third quantization, leading to a picture
where the problem can be collaterally studied in a reduced space that as such
provides a significant reduction in simulation costs. The purpose of this
manuscript is twofold, on the one hand it is to present a numerical method that
complements the Prosen's formalism by providing a protocol that efficiently
computes the system's NESS in tensorial representation in an exact way.  On the
other hand, this study intents to show how the aforementioned method has been
applied to determine the presence of uncoupled Majorana fermions in Kitaev chains
subject to baths on both ends using the criterion proposed in reference
\cite{reslen5}. The resulting phase diagram displays opposing features with respect
to the equilibrium map as well as coincidence over regions of parameter space
determined more by the hopping intensity than the chemical potential.  This paper
is divided as follows, section \ref{tous} describes how the third quantization
scheme has been implemented here and how the problem is reformulated from this
perspective. Section \ref{poly} shows how the NESS can be written as a product of
sums of Majorana fermions in operator space. In section \ref{tocca} the resulting
expression is decomposed as a product of next-site unitary operations acting on a
Fock state.  Key aspects of the numerical implementation are then discussed in
subsection \ref{fonandes}, while comparative simulations showing the error produced
by the proposed protocol are shown in subsection \ref{ericberg}. Section
\ref{pollock} documents the results obtained when the developed methods are applied
on an open Majorana chain with baths on the ends. Conclusions and final remarks are
finally presented in section \ref{besudo}.
\section{Migration to a second Fock space}
\label{tous}
Let us associate a string of ordered Majorana operators, denoted by $\hat{\tau}$, 
with a basis element of a fermion Fock space, written as $|\phi )$, in this way
\begin{gather}
\hat{\tau} = \hat{\gamma}_1^{n_1} ... \hat{\gamma}_{2 j -1}^{n_{2 j - 1}} \left(i \hat{\gamma}_{2 j} \right )^{n_{2 j}}  ... \left(i\hat{\gamma}_{2 N} \right)^{n_{2 N}} \Leftrightarrow  \nonumber \\ 
	\left | n_1...n_{2 j -1} n_{2 j}...n_{2 N} \right ) = |\phi).
	\label{thehand}
\end{gather}
The association is operational rather than physical since operators are being
associated with states instead of associating operators with operators or states
with states. The curved ket on the right serves as a remainder that the
corresponding Fock space is different from the original space of real fermions.
The inner product attached to the new Hilbert space satisfies the following
identity
\begin{gather}
Tr( \hat{\tau}'^\dagger \hat{\tau})  =  2^N ( \phi' | \phi ).  
\label{gnome}
\end{gather}
The equivalence can be extended to a superposition of objects since the expression 
is linear on both sides. Now consider the product $\hat{\gamma}_{2j-1} \hat{\tau}$. 
When $n_{2j-1}=0$, the result is \cite{schwabl}
{\small
\begin{gather}
(-1)^{\sum_{k=1}^{2j-2} n_k}   \hat{\gamma}_1^{n_1} ... \hat{\gamma}_{2 j -1} \left(i \hat{\gamma}_{2 j} \right )^{n_{2 j}}  ... \left(i\hat{\gamma}_{2 N} \right)^{n_{2 N}} \Leftrightarrow \tilde{c}_{2j-1}^\dagger \left | \phi \right ), \nonumber
\end{gather}
}
likewise, when $n_{2j-1}=1$ it is
{\small
\begin{gather}
(-1)^{\sum_{k=1}^{2j-2} n_k}   \hat{\gamma}_1^{n_1} ... \hat{\gamma}_{2 j -1}^0 \left(i \hat{\gamma}_{2 j} \right )^{n_{2 j}}  ... \left(i\hat{\gamma}_{2 N} \right)^{n_{2 N}} \Leftrightarrow \tilde{c}_{2j-1} \left | \phi \right ). \nonumber
\end{gather}
}
A single compact expression covering both cases reads
\begin{gather}
\hat{\gamma}_{2j-1} \hat{\tau} \Leftrightarrow \left( \tilde{c}_{2 j - 1} + \tilde{c}_{2 j - 1}^\dagger \right) | \phi ).
\end{gather}
\begin{table}
\begin{gather}
\begin{array}{|c|c|} \hline
\text{First space} & \text{Second space} \\ \hline
\hat{\gamma}_{2 j - 1} \hat{\tau}  & \left( \tilde{c}_{2 j - 1} + \tilde{c}_{2 j - 1}^\dagger \right) | \phi ) \\ \hline
i \hat{\gamma}_{2 j}  \hat{\tau}  & \left( -\tilde{c}_{2 j} + \tilde{c}_{2 j}^\dagger \right) | \phi ) \\ \hline
\hat{\tau}  \hat{\gamma}_{2 j - 1}  & \left( -\tilde{c}_{2 j - 1} + \tilde{c}_{2 j - 1}^\dagger \right) (-1)^{\tilde M} | \phi ) \\ \hline
\hat{\tau}  i \hat{\gamma}_{2 j}  & \left( \tilde{c}_{2 j} + \tilde{c}_{2 j}^\dagger \right) (-1)^{\tilde M} | \phi ) \\ \hline
\end{array} \nonumber
\end{gather}
\caption{Both left and right multiplication of a string of ordered Majorana operators by another operator have an equivalence on a fermionic Fock space.} 
\label{peto}
\end{table}
Following a similar analysis the equivalences reported in table \ref{peto} can be
derived. Linearity guarantees that identical relations are valid for a
superposition. Notice that ladder operators with a tilde have been used above to
differentiate these, which are understood as elements of a {\it second space}, from
the original modes on physical or {\it first space}. For instance, $\tilde{M}$ in
table \ref{peto} is the number operator in the second space,
\begin{gather}
\tilde{M} = \sum_{j=1}^{2 N} \tilde{c}_j^{\dagger} \tilde{c}_j,
\label{agatha}
\end{gather}
as such, it is not related to the actual total number of fermions in the system.
Moreover, the density matrix can be written in the first space as
\begin{gather}
\hat{\rho} = \sum_{n_1...n_{2N}} q_{n_1...n_{2N}} \hat{\gamma}_1^{n_1} ... \left(i\hat{\gamma}_{2 N} \right)^{n_{2 N}},
\end{gather}
the $q_{n_1...n_{2N}}$ being complex coefficients in general. In the second space 
the same concept goes over to
\begin{gather}
| {\rho} ) = \sum_{n_1...n_{2N}} q_{n_1...n_{2N}} | n_1...n_{2 N} ). 
\end{gather}
Because the parity operator in the second space, defined as $(-1)^{\tilde M}$,
commutes with $\tilde{\mathscr{L}}$, $|\rho)$ has a definite parity. The subsequent
development is designed for density matrices of even parity since this case covers
all instances of physical significance. Normalization requires
%
%
%{\footnotesize
\begin{gather}
tr(\hat{\rho}) = tr( \hat{\gamma}_1^0 ... \left(-i\hat{\gamma}_{2 N}\right)^0  \hat{\rho}) = 2^N (0...0 | \rho ) = 1.
\end{gather}
%}
%
%
Using the equivalences of table \ref{peto} and equations (\ref{ostinato}) and 
(\ref{baths}) it can be shown that the Lindblad equation (\ref{lindblad}) in the 
second space is given by
\begin{gather}
\frac{d |\rho ) }{d t}  =   \tilde{\mathscr{L}} | \rho ).
\end{gather}
Operator $\tilde{\mathscr{L}}$, which plays the role of a Liouvillian, comes to be
(valid for configurations of even parity)
{\scriptsize
\begin{gather}
\tilde{\mathscr{L}} = i\sum_{j=1}^{N} \sum_{k=1}^{N}  A_{2 j - 1,2 k} \left ( \tilde{c}_{2 k}^\dagger \tilde{c}_{2 j - 1} + \tilde{c}_{2 j - 1}^\dagger \tilde{c}_{2 k}\right ) + 2 \sum_n \nonumber \\ 
	  (-B_{2j-1}^{(n)} \tilde{c}_{2j-1}^\dagger + B_{2j}^{(n)} \tilde{c}_{2j}^\dagger)(B_{2k-1}^{(n)}  (\tilde{c}_{2k-1} + \tilde{c}_{2k-1}^\dagger ) + B_{2k}^{(n)} (-\tilde{c}_{2k} + \tilde{c}_{2k}^\dagger)) \nonumber \\
+  (B_{2j-1}^{(n)} \tilde{c}_{2j-1}^\dagger + B_{2j}^{(n)} \tilde{c}_{2j}^\dagger)(B_{2k-1}^{(n)}(-\tilde{c}_{2k-1} + \tilde{c}_{2k-1}^\dagger ) - B_{2k}^{(n)} (\tilde{c}_{2k} + \tilde{c}_{2k}^\dagger)).
\label{perales}
\end{gather}
}
Notice $\tilde{\mathscr{L}}$ is neither hermitian nor antihermitian. From a direct
substitution it can be proved that a totally occupied state is a right eigenstate of 
the Liouvillian
{\footnotesize
\begin{gather}
\tilde{\mathscr{L}} \left |  11...11  \right ) = L | 11...11 ), \text{ } L = -4 \sum_n \sum_j {B_{2j-1}^{(n)}}^2 + {B_{2j}^{(n)}}^2.
\end{gather}
}
The NESS in the second space satisfies
\begin{gather}
\tilde{\mathscr{L}} | N_{ESS} ) = 0.
\label{plataforma}
\end{gather}
Employing a second set of Majorana operators
\begin{gather}
\tilde{\gamma}_{2l-1} = \tilde{c}_l + \tilde{c}_l^\dagger \text{,  } \tilde{\gamma}_{2l} = i(-\tilde{c}_l + \tilde{c}_l^\dagger),
\label{cartier}
\end{gather}
the Liouvillian can be written as
\begin{gather}
\tilde{\mathscr{L}} = \sum_{j=1}^{4N} \sum_{k=1}^{4N} \mathscr{L}_{j k} \tilde{\gamma}_j \tilde{\gamma}_k.
\label{gymnopedies}
\end{gather}
Since the change of indexes $j \leftrightarrow k$ is essentially a cosmetic one, 
the Liouvillian coefficients must fulfill $\mathscr{L}_{j k} = -\mathscr{L}_{k j}$,
except when $j=k$, since diagonal elements can be finite in general and there is 
no reason to argue that the sum of diagonal coefficients is zero. The explicit form
of $\mathscr{L}$ can be consulted in appendix \ref{appendix1}.
\section{Obtention of the non-equilibrium stationary state}
\label{poly}
The NESS is calculated in second space via
\begin{gather}
| N_{ESS} ) = \lim_{t \rightarrow \infty}  e^{t \tilde{\mathscr{L}}} | 00\dots 0 ).
\label{crypto}
\end{gather}
This operation amounts to evolve a totally mixed density matrix over infinity time.
In this expression a normalization constant has been dropped because it cancels out
with the inner product constant of equation (\ref{gnome}) in all relevant
calculations of this work. Equation (\ref{crypto}) shows the NESS's parity is even
because it results as the evolution generated by a parity-preserving Liouvillian
applied over an even configuration. Equation (\ref{crypto}) is equivalent to
{\footnotesize
\begin{gather}
\lim_{t \rightarrow \infty} e^{t \tilde{\mathscr{L}}} \tilde{c}_{2N} \tilde{c}_{2N-1} \dots \tilde{c}_2 \tilde{c}_1| 11\dots 1 ) = 
\lim_{t \rightarrow \infty} e^{t \tilde{\mathscr{L}}} \prod_{j=2N}^{1} \tilde{c}_j | 11\dots 1 ), \nonumber
\end{gather}
}
which can also be written as
\begin{gather}
\lim_{t \rightarrow \infty} \left \{ \prod_{j=2N}^{1} e^{t \tilde{\mathscr{L}}} \tilde{c}_j e^{-t \tilde{\mathscr{L}}} \right \} e^{t \tilde{\mathscr{L}}} | 11\dots 1 ) = \nonumber \\ 
	\lim_{t \rightarrow \infty} e^{- t |L| } \prod_{j=2N}^{1} e^{t \tilde{\mathscr{L}}} \tilde{c}_j e^{-t \tilde{\mathscr{L}}} | 11\dots 1 ).
\label{chuby}
\end{gather}
Writing the modes in terms of (second) Majorana operators yields
{\small
\begin{gather}
\lim_{t \rightarrow \infty} e^{- t |L| } \prod_{j=2N}^{1} e^{t \tilde{\mathscr{L}}} \left( \frac{\tilde{\gamma}_{2j-1} + i\tilde{\gamma}_{2j}}{2} \right) e^{-t \tilde{\mathscr{L}}} | 11\dots 1 ).
\label{shigeru}
\end{gather}
}
Let us define evolved operators thus
\begin{gather}
\tilde{\gamma}_l(t) = e^{t \tilde{\mathscr{L}}} \tilde{\gamma}_l e^{-t \tilde{\mathscr{L}}}.
	\label{adastra}
\end{gather}
In this expression the contribution of diagonal elements in the Liouvillian cancels
out. Hence it is valid to make $\mathscr{L}_{jj}=0$ in (\ref{gymnopedies}) {\it
from now on}.  This does not mean that diagonal elements do not affect the NESS,
what happens is that such a contribution has been encapsulated in the overall
exponential factor of equation (\ref{shigeru}). Differentiation of equation
(\ref{adastra}) yields
\begin{gather}
\partial_t \tilde{\gamma}_l(t) = e^{t \tilde{\mathscr{L}}} [\tilde{\mathscr{L}},\tilde{\gamma}_l]  e^{-t \tilde{\mathscr{L}}}= -4 \sum_j \mathscr{L}_{l j} \tilde{\gamma}_j(t).
\label{apatros}
\end{gather}
Together with the initial condition, $\tilde{\gamma}_j(t=0) = \tilde{\gamma}_j$, 
this equation defines a solvable set of identities whose solution is given by
\begin{gather}
\tilde{\gamma}_l(t) = \sum_{j} e^{t z_j} Z_{l j} \tilde{q}_{j}. 
\label{salsakids}
\end{gather}
The unknown coefficients, $z_j$ and $Z_{l j}$, correspond to eigenvalues and 
right eigenvectors defined in the next manner
\begin{gather}
-4 \sum_{l} \mathscr{L}_{k l} Z_{l j} = z_j Z_{k j}.
\label{paco}
\end{gather}
The unknown operators, $\tilde{q}_j$, can be found from the initial condition
\begin{gather}
\tilde{\gamma}_l = \sum_{j} Z_{l j} \tilde{q}_{j} \rightarrow \tilde{q}_j = \sum_{l} Z_{j l}^{-1} \tilde{\gamma}_{l}.
\end{gather}
Replacing in equation (\ref{salsakids}) produces
\begin{gather}
\tilde{\gamma}_l(t) = \sum_{j=1}^{4N} \sum_{k=1}^{4N} e^{t z_j} Z_{l j} Z_{j k}^{-1} \tilde{\gamma}_{k}.
\end{gather}
As can be seen, the evolved operators are written in terms of the original
Majoranas. For $t$ finite the product in equation (\ref{shigeru}) is made up of
sums of such Majoranas and so can be expanded. Assuming that a NESS does exist and
is unique \cite{tomas_prosen}, terms of this expansion scaling slower that $e^{t
|L| }$ must vanish when $t\rightarrow \infty$, because of the overall exponential
term in (\ref{shigeru}).  Based on this observation, only contributions from
eigenvalues whose real parts add up to $|L|$ are kept in equation
(\ref{salsakids}). The set of these eigenvalues coincide the set of $z_js$ with
positive real part. Because the NESS is time independent, the remaining expression
must deliver the NESS for any value of $t$, making the actual value of $t$
irrelevant. Hence, time is  set to $t=0$.  Accordingly, an evolved operator
$\tilde{\gamma}_l(\infty)$ is replaced by
\begin{gather}
\tilde{s}_l = \sum_{k=1}^{4N} \sum_{j}  Z_{l j}  Z_{j k}^{-1} \tilde{\gamma}_{k} = \sum_{k=1}^{4N} S_{l,k} \tilde{\gamma}_k,
	\label{journey}
\end{gather}
in such a way that the sum over $j$ in the middle term includes only coefficients 
corresponding to eigenvalues $z_j$ with positive real part. Using these operators 
the state can be assembled as
\begin{gather}
|NESS ) = \prod_{j=2N}^{1} \left ( \frac{\tilde{s}_{2j-1} + i\tilde{s}_{2j}}{2} \right) | 11\dots 1 ) =  \nonumber \\
	\prod_{j=2N}^{1} \left ( \sum_{k=1}^{4N} R_{j,k} \tilde{\gamma}_k  \right) | 11\dots 1 ) = \prod_{j=2N}^{1} \tilde{f}_j | 11\dots 1 ),
	\label{nickjr}
\end{gather}
being $R_{j,k}$ time independent coefficients  that depend directly on the 
$S_{l,k}$ of equation (\ref{journey}).
\section{Folding of a complex stack}
\label{tocca}
The relation between the $\tilde{f}_j$s and $\tilde{\gamma}_k$s in equation 
(\ref{nickjr}) can be represented in matrix form whereupon both sets of operators 
are connected through a transfer matrix,
{\scriptsize
\begin{gather}
\left [
\begin{array}{c}
\tilde{f}_1    \\
\tilde{f}_2    \\
\vdots         \\
\tilde{f}_{2N}    
\end{array}
\right ] =
\left [
\begin{array}{ccccc}
R_{1,1} & R_{1,2}  & \dots & R_{1,4N-1}  & R_{1,4N} \\
R_{2,1} & R_{2,2}  & \dots & R_{2,4N-1}  & R_{2,4N} \\
\vdots    & \vdots       &       &  \vdots &  \vdots \\
R_{2N,1}& R_{2N,2} & \dots & R_{2N,4N-1} & R_{2N,4N} 
\end{array} 
\right ]
\left [
\begin{array}{c}
\tilde{\gamma}_{1}    \\
\tilde{\gamma}_{2}    \\
\vdots      \\
\tilde{\gamma}_{4N-1} \\   
\tilde{\gamma}_{4N} 
\end{array}
\right ]. 
\label{star}
\end{gather}
}
Here the right side of this equation is referred to as ``the stack'', in order to
emphasise a vertical ordering of sums of operators. In the traditional approach,
the solution process involves diagonalizing the transfer matrix all at once.  An
alternative is to work out the spectrum in layers of reductions, where on each
layer a single mode is decoupled until the problem is diagonal in some practical
sense.  Initially, let us point out that the coefficients can be complex and as
such the $\tilde{f}_j$s are not Majorana fermions in general. Neither are they
standard fermions because the Liouvillian transformation is not unitary.
Nevertheless,  anticommnutation rules prevail, 
\begin{gather}
	\{\tilde{f}_j,\tilde{f}_k\} = \lim_{t\rightarrow \infty}  e^{t \tilde{\mathscr{L}}}  \{\tilde{c}_j,\tilde{c}_k\} e^{-t \tilde{\mathscr{L}}} =  0. 
\end{gather}
This implies the coefficients display a relation somehow resembling orthogonality
\begin{gather}
\sum_l R_{j,l} R_{k,l} = 0.
\label{orto}
\end{gather}
It can be seen that this relation is invariant under similarity transformations.
The goal is to reduce (or fold) the transfer matrix using next-site unitary
operations in accordance with the strategy followed in reference \cite{reslen5} for
a matrix with real coefficients and orthogonal rows.  Neither of these conditions
are essential to fold the stack as shown forward.  The complication that arises
with complex coefficients is that they must be stripped of their complex phases
before any reduction can be implemented. To appreciate this point, let us see how a
standard phase transformation acts on a given Majorana operator
\begin{gather}
e^{i \varphi \hat{c}_j^{\dagger} \hat{c}_j } \tilde{\gamma}_{2 j - 1} e^{-i \varphi \hat{c}_j^{\dagger} \hat{c}_j }= e^{-i \varphi} \tilde{c}_j + e^{i \varphi} \tilde{c}^{\dagger}_j. \nonumber
\end{gather}
Hence, because the phases of $\hat{c}_j$ and $\hat{c}_j^\dagger$ spin in opposite
directions, the overall phase of $\tilde{\gamma}_{2 j - 1}$ cannot be shifted via a
local unitary operation. The reduction protocol being introduced, consists in
applying a series of next-neighbor unitary transformations over the NESS given by
equation (\ref{nickjr}) in order to simplify the transfer matrix (\ref{star}),
since changes induced over the state can be visualized as changes on the columns of
the transfer matrix. Having completed the reduction, the state can be recovered as
the inverse operation, which can be implemented numerically using the theory of
tensor product states. The reduction protocol can be summarized as follows 
\begin{enumerate}
\item Implement  
\begin{gather}
\tilde{U}_{1,4N} = e^{\frac{\theta_1}{2} \tilde{\gamma}_{4N} \tilde{\gamma}_{4N-1}}.
\label{mercedes}
\end{gather}
The scope of such a transformation is reduced to the modes 
involved therein, thus
\begin{gather}
\tilde{U}_{1,4N} (R_{4N-1,1} \tilde{\gamma}_{4N-1} + R_{4N,1} \tilde{\gamma}_{4N}) \tilde{U}_{1,4N}^{-1} = \nonumber \\
R_{4N-1,1}' \tilde{\gamma}_{4N-1} + R_{4N,1}' \tilde{\gamma}_{4N},
\label{lemale}
\end{gather}
where 
\begin{gather}
R_{1,4N-1}' = R_{1,4N-1} \cos \theta_1 + R_{1,4N} \sin \theta_1, \\
R_{1,4N}'= R_{1,4N} \cos \theta_1 - R_{1,4N-1} \sin \theta_1.
\end{gather}
The angle is chosen so as to make $Im(R_{1,4N}')=0$, which can be achieved by 
setting
\begin{gather}
\tan \theta_1 = \frac{Im(R_{1,4N})}{Im(R_{1,4N-1})}.
\end{gather}
Additionally, it is always possible to further gauge the angle to make 
$Im(R_{1,4N-1}')>0$. As a result the transfer matrix takes the form
\begin{gather}
\left [
\begin{array}{ccccc}
 \dots  & R_{1,4N-2}  & R_{1,4N-1}'  &    r_{1,4N}'         \\
 \dots  & R_{2,4N-2}  & R_{2,4N-1}'  & R_{2,4N}'   \\
        &  \vdots     &  \vdots      & \vdots        \\
 \dots  & R_{2N,4N-2} & R_{2N,4N-1}' &  R_{2N,4N}' 
\end{array} 
\right ],
\end{gather}
such that $r_{1,4N}' = Re(R_{1,4N}')$.	
\item A similar operation is applied with the intention of producing an analogous 
effect on the next pair of coefficients, like follows
\begin{gather}
\tilde{U}_{1,4N-1} = e^{\frac{\theta_2}{2} \tilde{\gamma}_{4N-1} \tilde{\gamma}_{4N-2}}.
\end{gather}
In accordance, the angle is set so that the imaginary part of $R_{1,4N-1}$ 
vanishes,
\begin{gather}
\tan \theta_2 = \frac{Im(R_{1,4N-1})}{Im(R_{1,4N-2})}.
\end{gather}
The transfer matrix would then look as
\begin{gather}
\left [
\begin{array}{ccccc}
 \dots  & R_{1,4N-2}'  & r_{1,4N-1}''  &    r_{1,4N}' \\
 \dots  & R_{2,4N-2}'  & R_{2,4N-1}''  & R_{2,4N}'    \\
        &  \vdots     &  \vdots      & \vdots         \\
 \dots  & R_{2N,4N-2}' & R_{2N,4N-1}'' &  R_{2N,4N}' 
\end{array} 
\right ].
\end{gather}
\item The process goes on, until all the coefficients but the first are made real.
\begin{gather}
\left [
\begin{array}{ccccc}
R_{1,1}'  & r_{1,2}''  & \dots  & r_{1,4N}'  \\
R_{2,1}'  & R_{2,2}''  & \dots  & R_{2,4N}'  \\
 \vdots   &  \vdots    & \vdots &  \vdots     \\
R_{2N,1}' & R_{2N,2}'' & \dots  & R_{2N,4N}'
\end{array} 
\right ].
\end{gather}
\item A new round of transformations is applied, starting with
\begin{gather}
\tilde{V}_{1,4N} = e^{\frac{\phi}{2} \tilde{\gamma}_{4N} \tilde{\gamma}_{4N-1}}.
\label{wars}
\end{gather}
The effect of this  is similar to (\ref{lemale}), the only difference is that the
coefficients are now real. The angle is chosen in such a way that the factor of
$\tilde{\gamma}_{4N}$ is canceled, which can be accomplished by making
\begin{gather}
\tan \phi = \frac{r_{1,4N}'}{r_{1,4N-1}''}.
\end{gather}
As a consequence, the matrix adopts the shape (apostrophes intentionally dropped)
\begin{gather}
\left [
\begin{array}{ccccc}
 \dots  & r_{1,4N-2}  & r_{1,4N-1}  &    0        \\
 \dots  & R_{2,4N-2}  & R_{2,4N-1}  & R_{2,4N}    \\
        &  \vdots     &  \vdots     & \vdots      \\
 \dots  & R_{2N,4N-2} & R_{2N,4N-1} &  R_{2N,4N} 
\end{array} 
\right ].
\end{gather}
\item A similar transformation is applied on the next pair of coefficients, causing
the elimination of $r_{1,4N-1}$ and leaving
\begin{gather}
\left [
\begin{array}{ccccc}
 \dots  & r_{1,4N-2}  &      0      &    0        \\
 \dots  & R_{2,4N-2}  & R_{2,4N-1}  & R_{2,4N}    \\
        &  \vdots     &  \vdots     & \vdots      \\
 \dots  & R_{2N,4N-2} & R_{2N,4N-1} &  R_{2N,4N} 
\end{array} 
\right ].
\end{gather}
\item This cancellation can be repeated on the subsequent coefficients, except for 
the last pair on the left corner since $R_{1,1}$ may not be entirely real. As a 
result the transfer matrix is reduced to 
\begin{gather}
\left [
\begin{array}{ccccc}
R_{1,1}  & r_{1,2}  &    0     & \dots  &    0      \\
R_{2,1}  & R_{2,2}  & R_{2,3}  & \dots  & R_{2,4N}  \\
 \vdots  &  \vdots  &  \vdots  & \vdots &  \vdots   \\
R_{2N,1} & R_{2N,2} & R_{2N,3} & \dots  & R_{2N,4N}
\end{array} 
\right ].
\end{gather}
\item The remaining pair of coefficients must obey equation (\ref{orto})
for $j=k=1$, therefore
\begin{gather}
R_{1,1}^2 + r_{1,2}^2 = 0 \rightarrow R_{1,1} = i r_{1,2}.
\end{gather}
This imply that the sum of modes in the first row becomes
\begin{gather}
i r_{1,2} \tilde{\gamma}_{1} + r_{1,2} \tilde{\gamma}_{2} = i r_{1,2} (\tilde{\gamma}_{1} -i \tilde{\gamma}_{2}) = 2 i r_{1,2} \tilde{c}_1^{\dagger}. \nonumber
\end{gather}
Observing that coefficients from different rows must obey equation (\ref{orto}) as
well, it follows for the first pair of coefficients on the second row
\begin{gather}
i r_{1,2} R_{2,1}  + r_{1,2} R_{2,2} = 0  \rightarrow R_{2,2} = -i R_{2,1}. \nonumber
\end{gather}
Adding the corresponding modes yields
\begin{gather}
R_{2,1} \tilde{\gamma}_{1} + R_{2,2} \tilde{\gamma}_{2} = R_{2,1} ( \tilde{\gamma}_{1} -  i \tilde{\gamma}_{2}) = 2 R_{2,1} \tilde{c}_1^{\dagger}. \nonumber
\end{gather}
The same applies over every row below the second row. This means that the reduction
has effectively eliminated the contribution of $\tilde{c}_1$. In addition, because
fermionic modes are nilpotent, $(\tilde{c}_1^\dagger)^2=0$, they make no
contribution except when they act only once. Since in order to find the state one
must multiply all the rows in the transfer matrix, it is therefore valid to cancel
the first pair of coefficients everywhere except on the first row, regardless of
their actual value, thus leaving 
\begin{gather}
\left [
\begin{array}{ccccc}
i r_{1,2} & r_{1,2}  &    0     & \dots  &    0      \\
   0      &    0     & R_{2,3}  & \dots  & R_{2,4N}  \\
 \vdots   &  \vdots  &  \vdots  & \vdots &  \vdots   \\
   0      &    0     & R_{2N,3} & \dots  & R_{2N,4N}
\end{array} 
\right ].
\end{gather}
\item An analogous protocol is applied on every but the last row, taking care not 
to affect the rows that have already been reduced. After this the transfer matrix 
turns into
\begin{gather}
\left [
\begin{array}{ccccccc}
i r_{1,2} & r_{1,2}  &    0     &    0     & \dots  &    0      &    0      \\
   0      &    0     &i r_{2,4} & r_{2,4}  & \dots  &    0      &    0      \\
 \vdots   &  \vdots  &  \vdots  &  \vdots  & \vdots &  \vdots   &  \vdots   \\
   0      &    0     &    0     &   0      & \dots  & \pm i R_{2N,4N} & R_{2N,4N}
\end{array} 
\right ]. \nonumber
\end{gather}
In principle, the folding can leave a plus or minus sign as indicated above,
however, for all sets of parameters studied here the sign has always turned up
positive. Anyhow, a negative sign does not produce any structural change in the
folding protocol. A specific consequence of the plus sign is that only
contributions from creation operators remain in the stack.
\end{enumerate}
The set of all transformations can be orderly bundled to produce a single operation
hereafter called $\tilde{T}$. From equation (\ref{nickjr}) the NESS can then be 
written like 
\begin{gather}
	|N_{ESS} ) = (-1)^N \tilde{T}^{-1} \left \{ \prod_{j=2N}^{1} 2 i r_{j,2j} \tilde{c}_j^{\dagger} \right \}  \tilde{T} | 11\dots 1 ),
\label{eau}
\end{gather}
so as explicitly
\begin{gather}
\tilde{T} = \prod_{l=2N-1}^{1}  \prod_{m=2l+1}^{4N} \tilde{V}_{l,m} \prod_{k=2l}^{4N} \tilde{U}_{l,k},
\end{gather}
wherein
\begin{gather}
\tilde{U}_{k,l} = e^{\frac{\theta_{k,l}}{2} \tilde{\gamma}_{k} \tilde{\gamma}_{k-1}} \text{ and } \tilde{V}_{k,l} = e^{\frac{\phi_{k,l}}{2} \tilde{\gamma}_{k} \tilde{\gamma}_{k-1}}.
\end{gather}
Angles $\theta_{k,l}$ and $\phi_{k,l}$ are determined according to the reduction 
protocol explained before. The change of sign for odd N comes from the order in 
which the operators inside curved parentheses add fermions on a vacuum state.
\subsection{Tensorial representation}
\label{fonandes}
Expression (\ref{eau}) can be simplified by noticing that the application of
creation operators between curved parentheses kills every basis state with the
exception of $|00...0)$, which contributes a coefficient that together with other
product factors defines a single multiplicative scalar. The function of this scalar
is quite elementary: it ensures the state is normalized by making the coefficient
of $|00...0)$ equal to one, but this can be done simply by inspecting the
coefficient of $|00...0)$ in the un-normalized state and then dividing the state by
this coefficient. Therefore, equation (\ref{eau}) is effectively equivalent to
\begin{gather}
|N_{ESS} ) = z_0 \tilde{T}^{-1} | 11\dots 1 ),
\label{extrait}
\end{gather}
being $z_0$ the complex constant that normalizes the NESS in the aforementioned
way. A key aspect of this NESS is that it has been decomposed as a series of
next-neighbors transformations. This makes it possible to implement a formulation
in terms of a canonical tensorial representation in a way that is now to be
described.  In a first step the state $|11...1)$ is written in tensor notation. The
series of operations represented by $\tilde{T}^{-1}$ is then applied over such a
state. This can be done using the protocols available to update a tensor structure
under the action next-site unitary transformations, which can most of the time be
done efficiently depending on the amount of entanglement present on the structure.
A conceptual description of the algorithm employed in this study to update a tensor
network under next site unitary operations can be found in the first appendix of
reference \cite{reslen5}. For a review on the subject of tensor network states see
for example reference \cite{orus}. The resulting structure is inspected for the
coefficient of $|00...0)$, and $z_0$ becomes the inverse of such a coefficient.
The resulting network of tensors together with $z_0$ form a structure that can be
used to calculate the system's observables.
\subsection{Test simulations}
\label{ericberg}
\begin{figure}[H]
\begin{center}
\includegraphics[width=0.32\textwidth,angle=-90]{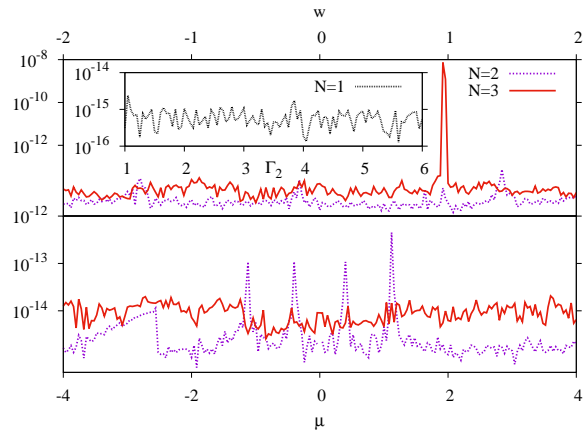}
\caption{Error, as calculated by equation (\ref{parfait}), 
estimating the NESS by the procedure described in the text. The case $N=1$  (inset)
is tested against expression (\ref{N1}) taking $\Gamma_1=1$. The cases
$N=2$ and $N=3$ are compared against the eigenstate with zero eigenvalue
of Liouvillian (\ref{perales}). The upper panel depicts the error for
$\mu=1$ and the lower one for $w=1.5$. In both cases $\Delta=1$, 
$\Gamma_{11}=1.3$, $\Gamma_{21}=2.2$, $\Gamma_{12}=3.4$ and $\Gamma_{22}=4.1$.} 
\label{fig1}
\end{center}
\end{figure}
In order to check the reliability of the proposal, the state obtained by the 
procedure recounted above has been compared against results extracted by other 
methods in a number of accessible instances. For the case $N=1$ analytical results 
can be derived considering the following baths operators
\begin{gather}
\hat{L}_1 = \sqrt{\Gamma_1} \hat{c} = \frac{\sqrt{\Gamma_1}}{2} \hat{\gamma}_1 + \frac{\sqrt{\Gamma_1}}{2} i \hat{\gamma}_2, \\
\hat{L}_2 = \sqrt{\Gamma_2} \hat{c}^\dagger = \frac{\sqrt{\Gamma_2}}{2} \hat{\gamma}_1 - \frac{\sqrt{\Gamma_2}}{2} i \hat{\gamma}_2.
\end{gather}
In this particular case the constants $w$ and $\Delta$ simply do not show up in the
Hamiltonian. Replacing (\ref{perales}) in (\ref{plataforma}) and solving yield
\begin{gather}
|N_{ESS}') = |00) + \frac{\Gamma_2 - \Gamma_1}{\Gamma_2 + \Gamma_1}|11).
\label{N1}
\end{gather}
In order to assess the difference against the state calculated by the folding 
procedure, $|N_{ESS})$, the following error estimate is introduced
\begin{gather}
\epsilon = \frac{||  | \psi  ) || }{|| |N_{ESS}') ||},
\label{parfait}
\end{gather}
where $|\psi) = |N_{ESS}) - |N_{ESS}')$. The inset in figure \ref{fig1} shows 
$\epsilon$ as a function of $\Gamma_2$ keeping $\Gamma_1$ constant. 

In order to test chains with $N=2$ and $N=3$ a couple of baths are added on both 
ends in the following fashion
\begin{gather}
\hat{L}_{1} = \sqrt{\Gamma_{11}} \hat{c}_1 \text{, } \hat{L}_{2} = \sqrt{\Gamma_{21}} \hat{c}_1^\dagger, \label{salsakids1} \\
\hat{L}_{3} = \sqrt{\Gamma_{12}} \hat{c}_N \text{, } \hat{L}_{4} = \sqrt{\Gamma_{22}} \hat{c}_N^\dagger. \label{salsakids2}
\end{gather}
Since in this instance there is no analytical solution, the NESS is calculated
numerically as the eigenstate of (\ref{perales}) associated with zero
eigenvalue.  Comparative errors can be seen in figure \ref{fig1}. As can be
observed, the protocol delivers the correct state for even- as well as odd-$N$
up to roundoff errors, which for the cases $N=2$ and $N=3$ might even come from
the benchmark calculation. Lack of analytical results for arbitrary $N$ makes
it difficult to analyze error scaling, but the study of a similar method on the
Kitaev chain reported in \cite{reslen5} indicates that relative errors saturate
for chains of some tens of sites to the order of magnitude of the square root
of machine precision.
\section{Results}
\label{pollock}
\begin{figure}[]
\begin{center}
\includegraphics[width=0.32\textwidth,angle=-90]{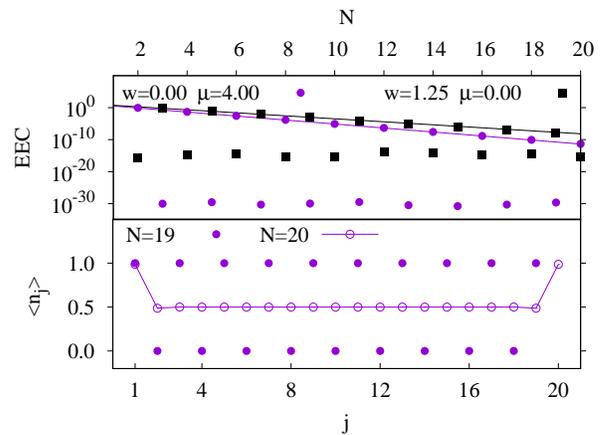}
\caption{Top. End-to-end correlations vs. chain size. When $w=0$, correlations
vanish for chains of odd size and decrease exponentially for chains of even
size.  The opposite happens when $\mu=0$. Bottom. Mean number of particles
vs position for $w=0$ and $\mu=4.00$. Hopping is suppressed and the NESS is
determined by the energy balance of occupied states. On chains of odd size
such a balance takes place when there is exactly one particle on every
other site, including the ends.  Such a state lacks any correlations.
Increasing the chain size by one breaks this order and provokes the charge
to disperse all over the chain interior, giving rise to EEC. This behavior
is characterized as a finite-size effect since its incidence on
correlations decays exponentially with $N$.  Everywhere in this figure
$\Delta=1$ and bath constants in equations (\ref{salsakids1}) and
(\ref{salsakids2}) are zero except $\Gamma_{21}=\Gamma_{22}=1$.}
\label{fig2}
\end{center}
\end{figure}
\begin{figure}[]
\begin{center}
\includegraphics[width=0.32\textwidth,angle=-90]{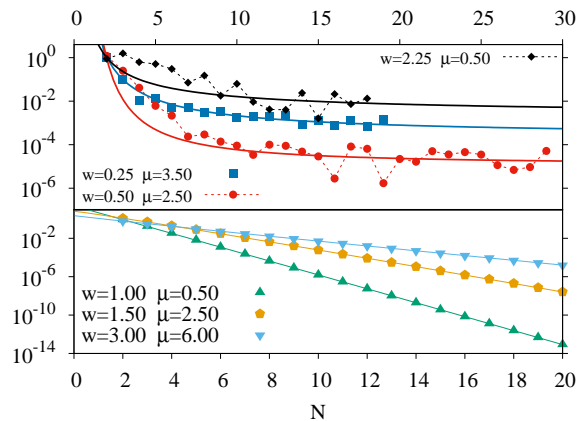}
\caption{End-to-end correlations vs. chain size. Top.  The decay pattern is
compatible with convergence to a finite value in the thermodynamic limit.
Finite size or boundary effects are notorious and do not seem to recede
over the ranges studied. This kind of behavior is indicated by orange
squares in the diagram of figure \ref{fig4}.  Bottom. Correlations decrease
exponentially as a function of $N$.  Finite-size effects are negligible.
This profile is demarked by blue circles in figure \ref{fig4}. Everywhere
nonzero constants are $\Gamma_{21}=\Gamma_{22}=\Delta=1$.}
\label{fig3}
\end{center}
\end{figure}
\begin{figure}[]
\begin{center}
\includegraphics[width=0.32\textwidth,angle=-90]{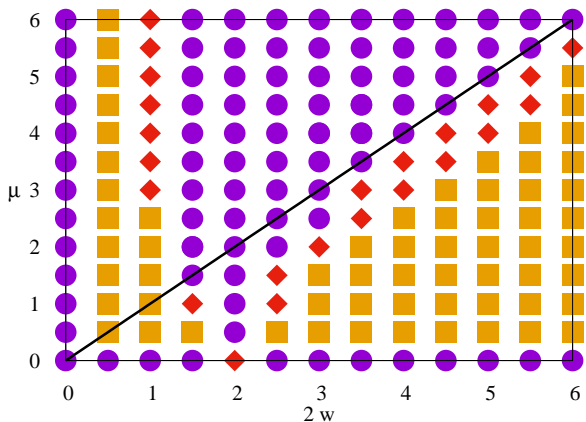}
\caption{Tentative phase diagram of a Kitaev chain with baths on the edges. Blue
circles signalize parameter sets where EECs vanish exponentially as
$N\rightarrow \infty$ in the way shown by figures \ref{fig2} or \ref{fig3}
(bottom), pointing to the absence of uncoupled Majorana fermions in the
infinite system.  Orange squares mark regions where scaling behavior
suggests the subsistence of correlations in the thermodynamic limit. Red
rhombuses indicate sets of parameters for which the tendency pattern is
unclear. The case $w=1$, $\mu=0$ is particular in that the corresponding
NESS seems to be non-unique because some $z_j$s in (\ref{paco}) are zero.
Nonzero constants are $\Gamma_{21}=\Gamma_{22}=\Delta=1$. The maximum chain
size lies in between $N=16$ and $N=30$ depending on the extend of
simulation time.  The black line allows to contrast with the equilibrium
case, where the topological phase $\mu < 2 w$ is known to display uncoupled
Majorana fermions. } 
\label{fig4}
\end{center}
\end{figure}
In solid state systems Majorana fermions are seen as collective excitations rather
than actual particles. Because the formalism assigns two Majoranas to every single
body state, it is common for Majorana fermions to couple with their twin mode,
giving in this way rise to localized excitations. However, it is possible that some
Majorana fermions did not pair, bringing about interesting phases that are highly
non-local and robust, also characterized as topological. The Kitaev chain is a
particular scenario where this behavior can be studied in detail due to the model
integrability. Such a possibility is however no longer an option in systems lacking
some level of analyticity. As an alternative, reference \cite{reslen5} introduces
an operational criterion that allows to determine whether the state contains
uncoupled Majorana fermions localized on the edges. The witness quantity is the
thermodynamic limit of the End-to-End Correlations (EEC) defined as
\begin{gather}
Z = \lim_{N\rightarrow \infty}	E_{EC} = \lim_{N\rightarrow \infty} 2 \left | \left \langle \hat{c}_1 \hat{c}_N^\dagger + \hat{c}_N \hat{c}_1^\dagger  \right \rangle \right |.
\label{pirri}
\end{gather}
The equivalent in the second space of the expression in brackets can be worked out 
as follows
\begin{gather}
\hat{c}_1 \hat{c}_N^\dagger + \hat{c}_N \hat{c}_1^\dagger =  i\hat{\gamma}_2 \hat{\gamma}_{2N-1} + \hat{\gamma}_1 i\hat{\gamma}_{2N} \nonumber \\ 
\Leftrightarrow |010...010) + |100...001) = |\Omega).
\end{gather}
Correlations can therefore be calculated as
\begin{gather}
E_{EC} = 2 |(\Omega |N_{ESS})|.
\end{gather}
The value of $Z$ can be estimated analyzing the behavior of EEC against growing 
$N$. However, given the difficulties in fitting some sets of data, in this work the
analysis is limited to determining whether or not $Z$ vanishes. Another observable 
of interest is the actual mean number of particles at a given position, which
comes from
\begin{gather}
\langle  \hat{n}_j \rangle = tr \left ( \hat{c}_j^\dagger \hat{c}_j  \hat{\rho} \right ) = \frac{1}{2} \left ( 1 + tr \left ( \hat{\gamma}_{2j-1} i \hat{\gamma}_{2j}  \hat{\rho} \right )  \right ).
\label{april161}
\end{gather}
Numerical simulations were carried out in Kitaev chains subject to baths described
by (\ref{salsakids1}) and (\ref{salsakids2}). The study has been limited to baths
with a particle-injection effect since this seems to be the most convenient
scenario to enhance EEC. Bath constants are therefore set to
$\Gamma_{21}=\Gamma_{22}=1$ and $\Gamma_{11}=\Gamma_{12}=0$.  The size of the
tensorial representation is a dynamical variable an depends on the requirements of
each particular computation. This explains why the parameter known as $\chi$ is not
reported. When $\chi$ is fixed there is a limit on the number of basis states
available to the system, and although this allows to obtain results for large
chains, it also affects the accuracy of the simulation, especially when long range
correlations are strong.

\begin{figure}[]
\begin{center}
\includegraphics[width=0.32\textwidth,angle=-90]{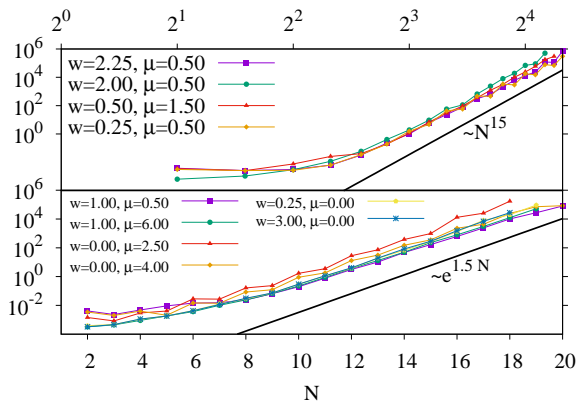}
\caption{Simulation time in seconds vs. chain size. Top. Time scaling is potential almost 
in general. Bottom. Scaling appears 
to be exponential in three particular cases: $w=0$, $w=1$ and $\mu=0$. Nonzero 
constants are $\Gamma_{21}=\Gamma_{22}=\Delta=1$.} 
\label{fig5}
\end{center}
\end{figure}

Let us initially address the results depicted in figure \ref{fig2}. It might appear
atypical that EEC can be nonzero in chains with zero hopping. On closer inspection
it is seen that chains of odd size display a separable particle distribution with
zero correlations, but in chains of even size an unstowed particle spreads all over
the chain interior and so enhances EEC. However, this correlations decay
exponentially with the chain size and make no contribution to $Z$. This latter fact
is characteristic of other configurations, for instance when $\mu=0$, although the
finite-size mechanism is different since in such a case hopping is nonzero.
Scanning over a grid of parameters it is possible to identify cases where, in
contrast, correlations seem to tend toward finite values, as shown in figure
\ref{fig3}. A trait that make it difficult to estimate $Z$ quantitatively is that
finite-size effects are strong and fitting attempts proved inconclusive.
Additionally, if finite-size effects are being enhanced by boundary effects, the
zig-zag pattern observed in the top pannel of figure \ref{fig3} might go on
nonstop, specifically in chains with long range correlations.  Regardless, the
observed dependency shows a pattern that is different from exponential decay.
Having established these two profiles, a potential phase diagram has been put
together in figure \ref{fig4}. The signs of $w$ or $\mu$ do not affect correlations
so that the diagram has been synthesized in a single quadrant. An useful benchmark
is the bathless chain, which is known to display uncoupled Majoranas in the region
$\mu < 2 w$. In this respect there seems to be coincidence for values of $w$
greater than $1.5$, discounting the line $\mu=0$. Apart from that, exponential
decay is seen along the whole line $w=1$, while convergence around finite values
can be seen in sectors where equilibrium states do not display uncoupled Majoranas,
like close to the $\mu$ axis. The opposite behavior, i.e., exponential decay in
sectors where the chain in equilibrium displays uncoupled Majoranas, takes place
over the $w$ axis and in some points close to the equilibrium boundary and the line
$w=1$.  In general terms it can be said that the system has in some degree resisted
the detrimental effects of the baths on its long range excitations, even has gained
in some sectors of the phase diagram, although the intensity of these excitations
has been negatively affected with respect to the equilibrium case.  Simulation
times are typically longer for chains with stronger EEC, but long times can be
displayed by chains with vanishing $Z$ as well, specifically along the lines $w=0$,
$w=1$ and $\mu=0$, as can be seen in the scaling profiles of figure \ref{fig5}.
Such scaling profiles reveal a gain in simulation efficiency for all sets of
parameters, even for those with exponential growth, because the problem dimension
scales exponentially with a slope that is no less than $2.8$. 
\section{Conclusions}
\label{besudo}
A method to compute the NESS of a general quadratic Fermi Hamiltonian under the
action of linear baths has been introduced and tested in a system known as the
Kitaev chain where a topological phase bearing uncoupled Majorana Fermions is well
characterized under equilibrium conditions. The protocol has been used to study the
incidence of uncoupled Majorana fermions in chains subject to baths at the ends
using the limit of correlations as a measure. The case in favor of uncoupled
Majoranas in some regions of parameter space is supported by the converging trends
of graphics of correlations vs. chain size. A tentative phase diagram has been
presented and contrasted with the equilibrium analogue showing more coincidence for
large values of the hopping constant.  Simulation times display potential scaling
against chain size for most sets of parameters, thus evidencing an improved
performance with respect to the scaling of the dimension of the original physical
space. 

The protocol introduced in this study has direct applications in a wide variety of
physical configurations of relevance. The second space approach of section
\ref{tous} is known as a super-fermion representation in the context of electron
transport \cite{kosov} and it is apparent that the methods introduced here suit
this field.  Changes can be incorporated to adjust the protocol in the direction of
the dependence of state with time. From this the net current flux
between the chain and the exterior could be calculated as the numerical derivative
of the total number of particles in the chain with respect to time. Another way of
studying transport phenomena is to add a tilted potential through a dependency of
$\mu$ with respect to position in (\ref{kita}) and then consider the current
through the chain as proportional to the mean value of the momentum operator. When
the fermion modes are sufficiently-localized Wannier-functions, such an operator
becomes a sum of next-neighbor hopping terms. Voltage would correspond to the slope
of the tilted potential.  In the same spirit, disorder effects can be studied by
assigning random coefficients to the chemical potential across the chain. None of
the aforementioned proposals would require any structural change in the method that
has been presented, which can be used as long as the Hamiltonian be quadratic in
the fermionic modes, and the bath terms be linear. The effect of lead contacts and
the calculation of zero-bias conductance could be addressed following the proposal
in reference \cite{Doornenbal}, which involves simulating the contacts as a set of
momentum modes coupled to the chain ends. Interaction with light can be studied
semiclasically without major changes to the current formulation. Spin systems can
be addressed via a Jordan-Wigner transformation. A more challenging project is to
develop an analogous formalism that worked with interacting systems.  This is
because in such a case it is not clear how to write the state as a product of sums
of operators. An option is to utilize expression (\ref{extrait}) as an ansatz.
Another option is to try to model interaction in a mean-field fashion. Also
relevant is the question of what are the minimum conditions a transfer matrix must
fulfill so that it can be folded in an applicable way. Similarly, the method looks
suitable to study quantities that involve the whole density matrix, as for example
mixedness or entropy, since the state is obtained in full.  Also of interest are
the insight possibilities that the method can offer to the field of fermionic
systems. In equation (\ref{extrait}) a fermion density-matrix is written as a
series of unitary operations. What this decomposition can provide in terms of
characterization of the physical state remains to be explored.  Thus far evidence
suggests that traces of uncoupled Majorana fermions and long range correlations can
be identified for specific sets of parameters in Kitaev chains exposed to baths on
the ends that break the mechanism of topological protection.  Moreover, the notion
of folding of modes can be used to study open quantum systems.

Financial support by Vicerrector\'ia de Investigaciones, Extensi\'on y Proyecci\'on
Social from Universidad del Atl\'antico is gratefully acknowledged.

\appendix
\section{Liouvillian coefficients in the second representation}
\label{appendix1}
Replacing the Majorana operators defined in equation (\ref{cartier}) and expanding, Liouvillian (\ref{perales})
becomes
{\small
\begin{gather}
\tilde{\mathscr{L}} =  \frac{1}{2} \sum_{j=1}^N \sum_{k=1}^N  A_{2j-1,2k} ( \tilde{\gamma}_{4k} \tilde{\gamma}_{4j-3} + \tilde{\gamma}_{4j-2} \tilde{\gamma}_{4k-1} ) + \sum_n \nonumber \\
- B_{2j}^{(n)} B_{2k}^{(n)} \tilde{\gamma}_{4j} \tilde{\gamma}_{4k} - B_{2j}^{(n)} B_{2k}^{(n)} \tilde{\gamma}_{4j-1} \tilde{\gamma}_{4k-1} + \nonumber \\
	- B_{2j-1}^{(n)} B_{2k-1}^{(n)} \tilde{\gamma}_{4j-2} \tilde{\gamma}_{4k-2} - B_{2j-1}^{(n)} B_{2k-1}^{(n)} \tilde{\gamma}_{4j-3} \tilde{\gamma}_{4k-3} + \nonumber \\
	2 i B_{2j-1}^{(n)} B_{2k}^{(n)} \tilde{\gamma}_{4j-3} \tilde{\gamma}_{4k} +  2 i B_{2j-1}^{(n)} B_{2k}^{(n)} \tilde{\gamma}_{4j-2} \tilde{\gamma}_{4k-1} + \nonumber \\
	2 i B_{2j-1}^{(n)} B_{2k-1}^{(n)} \tilde{\gamma}_{4j-2} \tilde{\gamma}_{4k-3} + 2 i B_{2j}^{(n)} B_{2k}^{(n)} \tilde{\gamma}_{4j} \tilde{\gamma}_{4k-1} + \nonumber \\ 
	2 B_{2j-1}^{(n)} B_{2k}^{(n)} \tilde{\gamma}_{4j-2} \tilde{\gamma}_{4k} + 2 B_{2j}^{(n)} B_{2k-1}^{(n)} \tilde{\gamma}_{4j-1} \tilde{\gamma}_{4k-3}.
\end{gather}
}
Although in this expression the Liouvillian coefficients $\tilde{\mathscr{L}}_{j k}$ do not form an antisymmetric
matrix, the anticommutation properties of Majorana fermions let us define conforming coefficients as follows
\begin{gather}
	\tilde{\mathscr{L}}_{j k}' = \frac{\tilde{\mathscr{L}}_{j k} - \tilde{\mathscr{L}}_{k j}}{2}, \text{ }   \tilde{\mathscr{L}}_{k j}' = -\tilde{\mathscr{L}}_{j k} \text{ for } j < k  \le 4 N.
\end{gather}
\end{document}